\documentclass[conference]{IEEEtran}
\IEEEoverridecommandlockouts
\usepackage{cite}
\usepackage{amsmath,amssymb,amsfonts}
\usepackage{algorithmic}
\usepackage{graphicx}
\usepackage{textcomp}
\usepackage{xcolor}
\usepackage{booktabs}
\usepackage{multirow}
\usepackage{graphicx}
\usepackage{makecell}
\usepackage{subfigure}
\def\BibTeX{{\rm B\kern-.05em{\sc i\kern-.025em b}\kern-.08em
    T\kern-.1667em\lower.7ex\hbox{E}\kern-.125emX}}
\begin{document}

\title{iAnomaly: A Toolkit for Generating Performance Anomaly Datasets in Edge-Cloud Integrated Computing Environments}

\author{\IEEEauthorblockN{Duneesha Fernando,
Maria A. Rodriguez and Rajkumar Buyya}
\IEEEauthorblockA{Cloud Computing and Distributed Systems (CLOUDS) Laboratory\\
School of Computing and Information Systems\\
The University of Melbourne, Australia\\
Email: dtfernando@student.unimelb.edu.au, \{marodriguez, rbuyya\}@unimelb.edu.au
}}

\maketitle
\thispagestyle{plain}
\pagestyle{plain}

\begin{abstract}
Microservice architectures are increasingly used to modularize IoT applications and deploy them in distributed and heterogeneous edge computing environments. Over time, these microservice-based IoT applications are susceptible to performance anomalies caused by resource hogging (e.g., CPU or memory), resource contention, etc., which can negatively impact their Quality of Service and violate their Service Level Agreements. Existing research on performance anomaly detection in edge computing environments is limited primarily due to the absence of publicly available edge performance anomaly datasets or due to the lack of accessibility of real edge setups to generate necessary data. To address this gap, we propose iAnomaly: a full-system emulator equipped with open-source tools and fully automated dataset generation capabilities to generate labeled normal and anomaly data based on user-defined configurations. We also release a performance anomaly dataset generated using iAnomaly, which captures performance data for several microservice-based IoT applications with heterogeneous QoS and resource requirements while introducing a variety of anomalies. This dataset effectively represents the characteristics found in real edge environments, and the anomalous data in the dataset adheres to the required standards of a high-quality performance anomaly dataset.
\end{abstract}


\section{Introduction}  
\label{sec:introduction}

Edge-cloud integrated environments consist of devices with heterogeneous computing, storage, and networking capabilities. Microservice architectures are increasingly used to modularize IoT applications and deploy them in these distributed environments to meet the Quality of Service (QoS) requirements of each module while optimizing resource usage \cite{al-doghman2023ai-enabled, WU2023Towards}. Over time, these microservice-based IoT applications are susceptible to performance anomalies caused by resource hogging (e.g., CPU or memory) and resource contention, which can negatively impact their QoS and violate their Service Level Agreements \cite{becker2020towards, Soualhia2019infrastructure, hunter2022deep}. Therefore, it is crucial to conduct performance anomaly detection on microservice-based IoT applications in edge computing environments and eventually mitigate such anomalies.

Currently, there is limited research on performance anomaly detection in edge computing environments. One of the main reasons for this is the absence of publicly available edge performance anomaly datasets, which are crucial for training and evaluating algorithms proposed in such research. The few existing studies rely on cloud datasets \cite{tuli2021generative, tuli2023aiaugmented} or data collected from private edge setups \cite{becker2020towards, Skaperas2024pragmatical, Soualhia2019infrastructure} to evaluate their proposed approaches. The cloud datasets have been collected from applications (mostly web applications) deployed on cloud servers. However, these cloud servers lack the heterogeneity found in edge devices in terms of computing, storage, and networking capabilities. Additionally, the microservices in cloud applications do not demonstrate the same diversity in terms of QoS and resource requirements as those in an IoT application. As a result, cloud datasets fail to capture characteristics inherent to real edge environments. On the other hand, private edge setups have not been publicly released and lack detailed information, which makes it difficult to replicate their environments, generate the necessary data, and reproduce the results of the anomaly detection experiments. It also hinders research in the field because not everyone has access to a real edge-cloud deployment for data collection purposes. Hence, relying on cloud datasets and private edge setups does not facilitate performance anomaly detection research in edge computing environments, thus posing a challenge to the progression of the field. Therefore, there is an opportunity to create a performance anomaly dataset that reflects the characteristics of edge computing environments and release the setup used for dataset generation. 

Edge computing emulators are a suitable platform to generate performance anomaly datasets. They are more representative of real edge environments when compared to simulators, and are more easily accessible and cost-effective when compared to real edge deployments. The main aim of existing edge computing emulators is to create a staging environment that achieves compute and network realism similar to a real edge environment and facilitate testing of IoT applications before deploying them into production \cite{mayer2017emufog, coutinho2018fogbed, hasenburg2019mockfog, Zeng2019emuedge, symeonides2020fogify, negin2024icontinuum}. However, these general-purpose emulators do not incorporate in their design, tools and mechanisms required to autonomously and transparently generate large-scale performance anomaly datasets useful for model training and evaluation. For example, they lack adequate monitoring tools to collect performance and system-level metrics, workload generation tools to generate and capture normal performance data, and chaos engineering mechanisms to inject performance anomalies into applications.  

This work addresses this gap by presenting the iAnomaly framework, a performance anomaly-enabled full-system emulator that accurately models an edge computing environment hosting microservice-based IoT applications. iAnomaly is designed with open-source tools and provides fully automated dataset generation capabilities to generate labeled normal (data collected under normal conditions without anomalies) and anomaly (data collected under anomalous conditions) data based on user-defined configurations. In addition, we present a performance anomaly dataset generated using the proposed framework. The dataset captures performance data for several microservice-based IoT applications with heterogeneous QoS and resource requirements across a wide range of domains, software architectures, service composition patterns, and communication protocols by introducing a variety of client/sensor-side as well as server-side anomalies. To the best of our knowledge, this multivariate dataset is the first open-source edge performance anomaly dataset.

The analysis of the dataset showed that the microservices within it vary in terms of their QoS and resource usage during regular operation, thus successfully capturing the characteristics of a real edge dataset. Further analysis confirmed that the anomalous data in the dataset meets the necessary standards for a high-quality performance anomaly dataset. This includes having an anomaly ratio comparable to other standard anomaly datasets and the dataset's non-triviality.

The rest of the paper is organized as follows: Section \ref{sec:related work} reviews the existing related works. Section \ref{sec:main_approach} presents the architecture of the iAnomaly toolkit, while section \ref{sec:implementation} discusses the implementation aspects of iAnomaly. Section \ref{sec:case_study} provides details of the generated performance anomaly dataset followed by an analysis of the dataset. Section \ref{sec:conclusion} concludes the paper and draws future research directions.

\section{Related Work}
\label{sec:related work}

Out of the existing research studies conducted around performance anomaly detection in edge computing environments, Becker et al. \cite{becker2020towards}, Soualhia et al. \cite{Soualhia2019infrastructure}, and Skaperas et al. \cite{Skaperas2024pragmatical} evaluated their proposed approaches using data collected from private edge setups. However, the lack of detailed information about these private edge setups makes it challenging to reproduce their environments and generate the necessary data to replicate the results of their anomaly detection experiments. Tuli et al. \cite{tuli2021generative} evaluated their proposed approaches using two publicly available cloud datasets: the Server Machine Dataset (SMD) collected from a large Internet company \cite{su2019robust}, and the Multi-source Distributed System (MSDS) dataset generated from microservices deployed on a cluster of bare metal nodes with homogeneous computing, storage, and network capabilities \cite{sasho2020multisource}. As a result, these cloud datasets are unable to accurately represent the properties inherent to real edge environments. Tuli et al. also conducted a further evaluation on three self-created datasets collected from a private edge setup. However, similar to the rest of the literature, they also do not provide sufficient details required for reproducing the data generation environments. Therefore, the reliance on cloud datasets and private edge setups presents a challenge to the progression of the field of performance anomaly detection research in edge computing environments. Additionally, there is a lack of publicly available normal traces from real edge environments into which anomalies can be injected to generate synthetic datasets. We further explore this gap in current research by creating and releasing a performance anomaly dataset that reflects the characteristics of edge computing environments along with the setup used for dataset generation.

There are three options of platforms for generating a performance anomaly dataset. They are simulators, emulators, and real edge environments. Out of these, emulators employ real applications deployed on testbed hardware to emulate real-world infrastructure configurations \cite{negin2024icontinuum}, while simulators do not support real-world IoT protocols and services \cite{coutinho2018fogbed}. Simulations make a number of simpliﬁcations that may not always hold true, especially with an infrastructure as dynamic as edge computing \cite{mayer2017emufog}. Most simulators lack detailed network simulation capabilities and focus on speciﬁc aspects of edge modeling, such as service scheduling \cite{negin2024icontinuum}. Therefore, we identify emulators as the most suitable platform for generating data as they provide a higher Degree of Realism (DoR) than simulators and because they enable the generation of large-scale data in a cost-effective manner as opposed to real edge environments. 

\begin{table*}[t]
\tiny
\caption{Comparison of performance anomaly dataset generation capabilities in edge emulators}
\label{table:relatedWork}
\resizebox{\textwidth}{!}{%
\begin{tabular}{|p{1.5cm}|p{3.5cm}|p{1cm}|p{1cm}|p{2cm}|p{1cm}|p{2.75cm}|}
\hline
\textbf{Edge emulator/ toolkit} &
  \textbf{Main objective of work} &
  \textbf{Fault injection capabilities} &
  \textbf{Performance anomaly dataset generation capabilities} &
  \textbf{Emulation capabilities/ architecture} &
  \textbf{Microservice support} &
  \textbf{Applications} \\ \hline
EmuFog \cite{mayer2017emufog} &
Testing IoT apps in a staging environment before deploying into the real edge, Automatic placement of fog nodes.
  &
  $\times$ &
  $\times$ &
  Focused on network emulation &
  $\times$ &
  Not reported \\ \hline
FogBed \cite{coutinho2018fogbed} &
 Creating an environment to conduct resource management/service orchestration experiments. &
  $\times$ &
  $\times$ &
  Focused on network emulation, Container-based emulator &
  $\times$ &
  Healthcare prevention and monitoring system \\ \hline
MockFog \cite{hasenburg2019mockfog} &
Testing IoT apps in a staging environment before deploying into the real edge. &
  $\times$ &
  $\times$ &
  Full-system emulator &
  $\times$ &
  Ambulance cars communicating vital measures to hospitals \\ \hline
EmuEdge \cite{Zeng2019emuedge} &
Achieving compute and network realism of a real edge environment. &
  $\times$ &
  $\times$ &
  Full-system emulator &
  $\times$ &
  Not reported \\ \hline
Fogify \cite{symeonides2020fogify} &
Testing IoT apps in a staging environment before deploying into the real edge. &
  \checkmark &
  $\times$ &
  Container-based emulator &
  \checkmark &
  Smart transport applications \\ \hline
iContinuum \cite{negin2024icontinuum} &
Achieving compute and network realism of a real edge environment, Intent-based emulation. &
  \checkmark &
  $\times$ &
  Full-system emulator &
  \checkmark &
  Image processing application \\ \hline
iAnomaly (proposed)&
Creating a toolkit to generate performance anomaly datasets. &
  \checkmark &
  \checkmark &
  Full-system emulator &
  \checkmark & Face detection/recognition application, Industrial machinery predictive maintenance application, Location retrieval application\\ \hline
\end{tabular}%
}
\end{table*}

Existing edge computing emulators can be organized under two main categories based on the level of virtualization and abstraction used to model the edge devices. They are 1) full-system emulators and 2) container-based emulators. In container-based emulators, edge devices are represented as docker containers, while full-system emulators provide a higher granularity of emulation by allowing the deployment of multiple containerized microservices within edge devices modeled as virtual machines (VMs). Early emulators such as EmuFog \cite{mayer2017emufog}, FogBed \cite{coutinho2018fogbed}, MockFog \cite{hasenburg2019mockfog}, and EmuEdge \cite{Zeng2019emuedge} do not support the microservice-level granularity of IoT applications. Fogify \cite{symeonides2020fogify} is the first edge emulator to support the microservice-level granularity of IoT applications. However, it is limited to deploying only a single microservice per edge device due to being a container-based emulator. Extending such emulators with dataset generation capabilities restricts data collection to device-level anomalies only. In contrast, iContinuum \cite{negin2024icontinuum} is a full-system emulator with support for microservices deployment. Unlike container-based emulators
, full-system emulators provide a higher level of realism and also allow the injection of both device-level and microservice-level anomalies. Consequently, our research aims to bridge the identified gap by developing a full-system emulator with performance anomaly dataset generation capabilities. A comparison of performance anomaly dataset generation capabilities in existing edge emulators \cite{mayer2017emufog, coutinho2018fogbed, hasenburg2019mockfog, Zeng2019emuedge, symeonides2020fogify, negin2024icontinuum} along with our proposed iAnomaly toolkit is shown in Table \ref{table:relatedWork}.   

The main intention of existing emulators is to test IoT applications in a staging environment before deploying them into production. They are designed to achieve the compute and network realism of an edge environment, and the evaluation of these studies is also focused on those aspects. Modern emulators such as Fogify \cite{symeonides2020fogify} and iContinuum \cite{negin2024icontinuum} also have the capability to perform fault injections. However, the main intention of such fault injection capabilities is not to collect data for performance anomaly detection model training but to test the fault tolerance and availability aspects of IoT applications in the face of faults.  

Although both Fogify and iContinuum have implemented and evaluated fault injection capabilities, neither of them has specified the tools used to inject anomalies. Since these emulators only conducted injections of a limited number of anomaly types, it can be inferred that they likely utilized basic tools, such as stress-ng, for this purpose. However, such mechanisms do not allow for the introduction of failures or disruptions in a controlled manner.

Both Fogify and iContinuum are capable of collecting both system and application metrics during monitoring. However, Fogify utilizes an in-house developed monitoring tool for this purpose, whereas open-source tools are preferred in emulators to support interoperability and transparency of code. iContinuum employs sFlow-RT, an open-source tool, to capture network and host-level metrics such as CPU and memory usage. It also integrates sFlow agents with Prometheus, another open-source tool, to collect application metrics. While Fogify requires explicit instrumentation, i.e. manually embedding monitoring code within the source code, in order to capture performance metrics, sFlowRT can only capture application metrics without explicit instrumentation from IoT applications that communicate via HTTP protocol. As most IoT applications deployed in edge computing environments are not limited to HTTP protocol and use a variety of protocols such as MQTT, RTSP, etc., it is important to be able to collect metrics from applications communicating via such non-HTTP protocols as well.

Fogify has not provided details of its workload generation tool, while iContinuum uses Locust for workload generation. However, Locust primarily focuses on generating HTTP/HTTPS workloads, while it is important to incorporate a workload generation tool supporting a wide range of protocols, not just HTTP. 

Consequently, it is evident that the current emulators lack the necessary tools for generating performance anomaly data. These tools include a monitoring tool for collecting metrics, a workload generation tool for creating normal performance data, and a chaos engineering tool for injecting performance anomalies. Identifying this research gap, our paper aims to address it by developing a toolkit with an emulator that incorporates a set of open-source tools for generating performance anomaly datasets. 

In addition to finding the best open-source tools for generating performance anomaly datasets and creating a full-system emulator with performance anomaly dataset generation capabilities, we also integrate automated dataset generation features into our proposed iAnomaly toolkit. Moreover, as shown in Table \ref{table:relatedWork}, most emulators have released only one IoT application which they used in their experiments. However, we generate (and release) an open-source performance anomaly dataset using iAnomaly by deploying three IoT applications consisting of microservices with varying QoS and resource requirements. These applications span a wide range of domains, software architectures, service composition patterns, and communication protocols.  

\section{iAnomaly Architecture}
\label{sec:main_approach}

\begin{figure}[t]
\centerline{\includegraphics[width=\columnwidth]{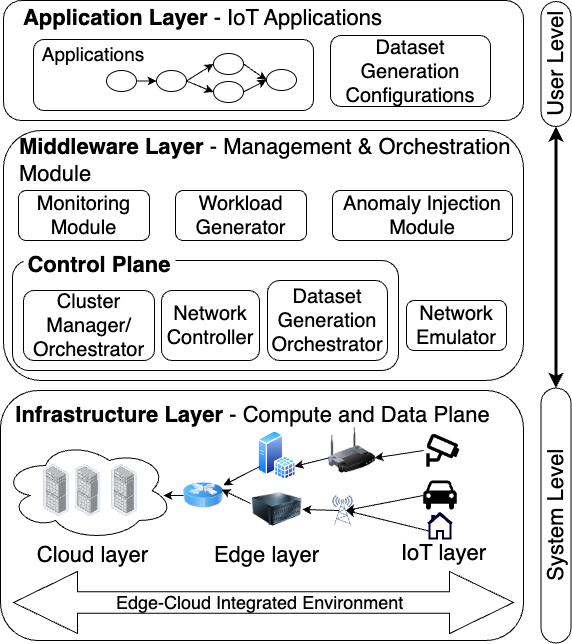}}
\caption{System architecture of iAnomaly}
\label{toolkit_architecture}
\vspace{-15pt}
\end{figure}

Figure \ref{toolkit_architecture} depicts the architecture of iAnomaly. At the core of the framework is a full system emulator that comprises multiple distinct layers with a set of components to build all the layers from infrastructure to applications. The infrastructure layer hosts a diverse array of computing and networking resources. While the heterogeneity of compute nodes can be emulated using Virtual Machines (VM) with different resource capacities in the cloud, a network emulator is used to construct the network topology (by creating virtual switches and network elements) and simulate the network within the infrastructure layer. Users can define the application structure of the microservice-based IoT applications through the application layer. The middleware layer of a full-system emulator manages the deployment and operation of applications across the emulated infrastructure. Its control plane consists of a cluster manager/orchestrator that utilizes containerization and orchestration technologies to manage the computing cluster and its resources, and a network controller that uses Software-Defined Networking (SDN) technologies to manage the network (such as effectively regulating the network flow while considering resource usage conditions). 

In addition to a full system emulator, iAnomaly includes components to facilitate the data generation and collection process within its middleware layer. These components consist of a monitoring module, a workload generation tool, and a tool for injecting anomalies.

\textbf{The monitoring module} is responsible for gathering system and application metrics from the deployed microservices. The collected data will be stored in a database and retrieved back through queries when creating the dataset. An important requirement of the monitoring tool is to be able to collect data from IoT applications communicating not only via HTTP-based protocols but also via non-HTTP protocols such as Kafka, MQTT, and RTSP. In addition, it is preferable for the tool to be capable of collecting application metrics from programs without the need for explicit instrumentation.

\textbf{The workload generator} is in charge of sending request/sensor data to the microservices. This tool is used during normal data generation as well as for introducing client/sensor-side anomalies such as user surges and spikes. Details of request workloads, including concurrency and duration, are specified using test plans. It is important that the workload generation tool also supports a wide range of protocols, not just HTTP. When generating data from a specific microservice, we need at least two instances of workload generators, one for generating normal workloads, and another for introducing client/sensor-side anomalies.

\textbf{The anomaly injection tool} is responsible for injecting server-side anomalies, such as resource hogging and service failures. MockFog \cite{hasenburg2019mockfog}, which is one of the early edge emulators, suggested using chaos engineering tools for injecting anomalies and conducting performance testing. Chaos engineering tools are designed to test the fault tolerance of systems by deliberately introducing failures or disruptions in a controlled manner, making them suitable for inclusion in the proposed toolkit to inject performance anomalies. Moreover, by incorporating chaos engineering tools, we can inject a diverse range of anomalies, unlike with basic tools such as stress-ng. 

Generating a significant amount of normal and anomaly data by using these data collection tools is a time-consuming and repetitive task necessitating human intervention. Additionally, there is a learning curve associated with using the tools, notably in terms of creating necessary test plans (incorporating varying parameters representing normal and anomalous workloads) using the workload generator, designing chaos engineering configurations, and scripting data collection for retrieving information from the monitoring tool. 

To overcome these challenges, we further extend the iAnomaly framework with automated dataset generation capabilities, where users can provide the configurations of the expected dataset, and the resulting labeled normal and anomaly data will be stored in a predefined location. As depicted in Figure \ref{toolkit_architecture}, users can define the dataset generation configurations through the application layer. We also introduce a dataset generation orchestrator to the middleware layer, which interprets the content from the dataset generation configuration and coordinates with the data generation tools in the toolkit to generate the normal and anomaly data required for the performance anomaly dataset. This process will be explained in detail in the next section.

\section{iAnomaly Implementation}
\label{sec:implementation}

\begin{figure}[t]
\centerline{\includegraphics[width=\columnwidth]{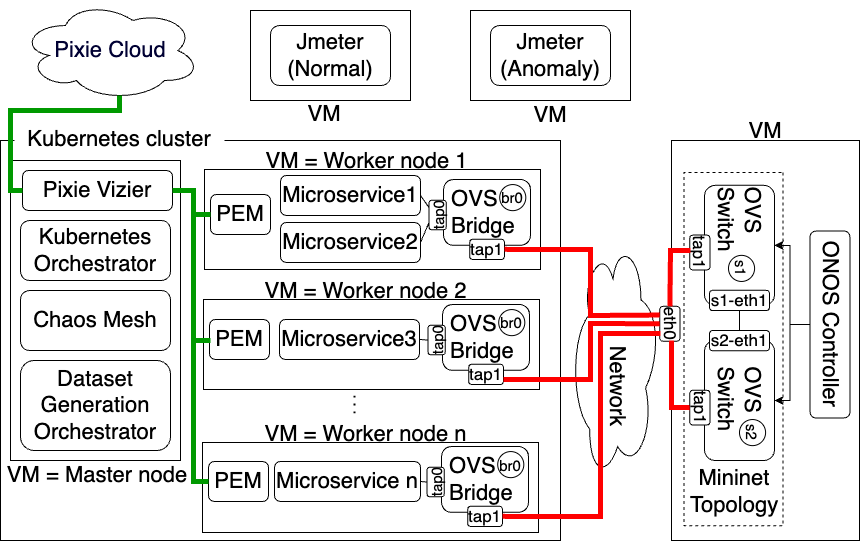}}
\caption{Deployment diagram of the iAnomaly toolkit}
\label{deployment_diagram}
\vspace{-15pt}
\end{figure}

This section describes the implementation details of the architecture presented in section \ref{sec:main_approach}. The deployment diagram of the iAnomaly toolkit is shown in Figure \ref{deployment_diagram}. iAnomaly relies on iContinuum \cite{negin2024icontinuum} as a full-system emulator to accurately model edge computing environments hosting microservice-based IoT applications. iContinuum utilizes Virtual Machines (VMs) with varying resource specifications to demonstrate the heterogeneity of edge devices in terms of computing and storage. Additionally, a VM with higher resource capacities acts as the master node, hosting only the tools required for compute orchestration. Acknowledging the critical role of the master node as the system’s control plane, we ensure that microservices are deployed only on VMs other than the master node. Following the implementation of iContinuum, iAnomaly uses Mininet\footnote{https://mininet.org/} as the network emulator to construct the network topology and simulate the bandwidth between edge devices. The OVS switches that form the Mininet topology are deployed in a separate VM, which is also nondeployable for microservices. 

In line with iContinuum's control plane implementation, iAnomaly also uses Kubernetes, specifically K3s\footnote{https://k3s.io/}, which is a lightweight Kubernetes distribution designed for resource-constrained environments such as edge computing or IoT devices, as the cluster manager/orchestrator. Additionally, iAnomaly utilizes Open Network Operating System (ONOS)\footnote{https://opennetworking.org/onos/} as the network controller, and it manages the OVS switches in the Mininet topology. The ONOS controller is also deployed in the VM where the OVS switches are deployed. Figure \ref{deployment_diagram} shows how the Kubernetes orchestrator is deployed in the master node and forms a multi-node Kubernetes cluster with the worker nodes. Each worker node is configured with an OVS bridge featuring two virtual interfaces, tap0, and tap1, out of which tap1, which is configured as a GRE interface, is linked to the tap1 port of the corresponding OVS switch. This ensures that the worker nodes have a bi-directional connection with the Mininet-created network topology through Generic Routing Encapsulation (GRE) tunnelling.

The Monitoring Module is realized by using Pixie\footnote{https://docs.px.dev/}, a lightweight and open-source eBPF(extended Berkeley Packet Filter)-based monitoring tool specifically designed for Kubernetes applications. eBPF-based monitoring tools, which gained popularity recently, allow sandboxed programs to execute directly inside the Linux kernel and automatically capture telemetry data in a non-intrusive manner, i.e., without requiring modifications to user-space applications \cite{Levin2020ViperProbe}. It also supports monitoring a wide variety of protocols, including HTTP, Kafka, AMQP, and MySQL, making it well-suited for monitoring edge computing environments. When Pixie is deployed via the toolkit, a Pixie Edge Module (PEM) is deployed for each worker node. These modules capture monitoring data from microservices deployments on the worker nodes and send those to the Pixie Vizier deployed on the Kubernetes master node, from where they are transferred to the Pixie cloud. Pixie Vizier acts as Pixie's central collector and is also responsible for managing PEMs. When retrieving back the collected data, data retrieval queries written in Pixie language (PxL) are executed against the Pixie cloud via a Pixie API client.

The Workload Generator is implemented by leveraging Jmeter\footnote{https://jmeter.apache.org/} as it supports a wide range of protocols, not just HTTP. We are using two separate instances of JMeter: one to generate normal workloads and the other to create client/sensor-side anomalies. Both instances are deployed outside the Kubernetes cluster to run the JMeter loads as needed and to avoid interfering with the master node's operations. 

Chaos Mesh\footnote{https://chaos-mesh.org/}, an open-source chaos engineering platform for Kubernetes, is used as the anomaly injection tool and deployed in the Kubernetes master node. From there, CRD (Custom Resource Definition) YAMLs are applied to introduce server-side anomalies into the target deployments. Chaos Mesh is capable of injecting a multitude of server-side anomalies such as CPU stress, memory stress, network delay, etc. 

\begin{figure}[t]
\centerline{\includegraphics[width=\columnwidth]{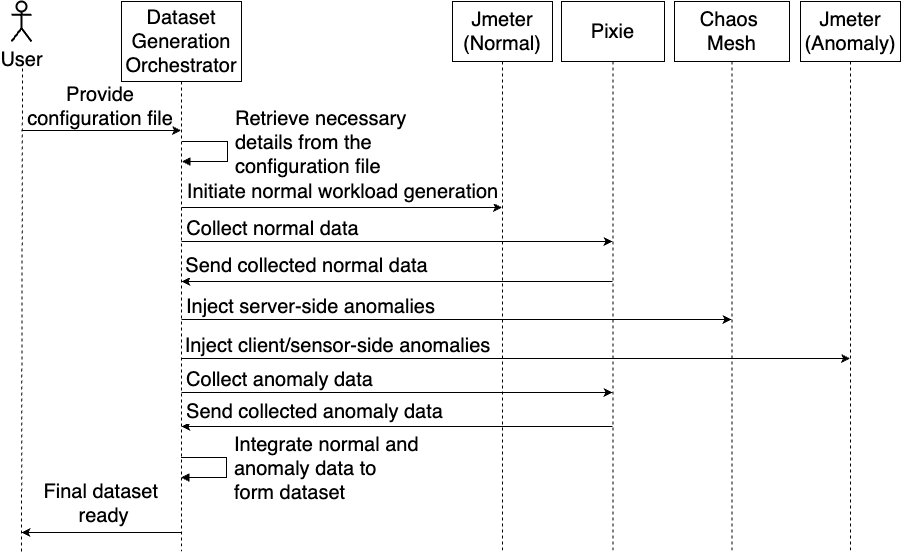}}
\caption{Interactions between the dataset generation orchestrator and other components}
\label{toolkit_components}
\vspace{-15pt}
\end{figure}

The dataset generation orchestrator is deployed in the Kubernetes master node. As illustrated in Figure \ref{toolkit_components}, it acts as the central component that interprets the content from the dataset generation configuration/s and coordinates with the data generation tools in the toolkit to produce the labeled normal and anomaly data needed for the performance anomaly dataset. To start the data generation process, the dataset generation orchestrator reads the configuration file/s to retrieve the deployment details, normal data collection parameters, and anomaly injection settings. It then remotely executes the test plans on Jmeter's normal data generation instance using Paramiko's\footnote{https://docs.paramiko.org/} SSH client to initiate normal workload generation. Once the normal data has been generated for the required duration, the orchestrator executes PxL queries to collect the generated normal data. Simultaneously, while the normal workload generation process is ongoing, the orchestrator proceeds to inject anomalies. Anomalies are injected either through server-side disruptions using Chaos Mesh or via client/sensor-side anomalies through Jmeter's anomalous data generation instance. For server-side disruptions, the dataset generation orchestrator applies the corresponding chaos YAMLs through the installation of helm charts\footnote{https://helm.sh/}. After the entire anomaly injection period, the orchestrator executes PxL queries to collect the corresponding anomaly data. Finally, both normal and anomaly data are funnelled back to the orchestrator, which integrates and processes the data to create a comprehensive dataset suitable for training and evaluating performance anomaly detection models.

Therefore, the realization of the architecture proposed in section \ref{sec:main_approach} using the open-source tools discussed earlier, has made the process of dataset generation easily accessible for researchers. We have released the source code of the iAnomaly toolkit, which includes iContinuum as its full-system emulator together with the chosen open-source data collection tools and the code for automated dataset generation, in a public repository\footnote{https://github.com/Cloudslab/iAnomaly}.

\section{Case Study: Dataset Generation}
\label{sec:case_study}

This section showcases how iAnomaly was used to create an open-source labeled dataset consisting of normal and anomalous data collected for three different IoT applications, followed by an analysis of the generated performance anomaly dataset.

\subsection{IoT Applications}
\label{subsec:iot_apps}

\begin{table*}[t]
\caption{Properties of IoT applications used to generate performance anomaly dataset}
\label{tab:iot-app-properties}
\resizebox{\textwidth}{!}{%
\begin{tabular}{|c|c|c|c|c|c|}
\hline
\textbf{Application} &
  \textbf{Microservice} &
  \textbf{Type of task performed} &
  \textbf{Software architecture} &
  \textbf{\makecell{Service \\composition \\pattern}} &
  \textbf{QoS properties} \\ \hline
\multirow{4}{*}{\shortstack{Face detection/\\ recognition application}} &
  Preprocessor &
  Computer vision-based &
  Stream processing &
  \multirow{4}{*}{Chained} & LC, HTp,HCI, BI
   \\ \cline{2-4} \cline{6-6} 
 & Face Detector            & Computer vision-based               & Request-response & & LC, MTp, HCI \\ \cline{2-4} \cline{6-6} 
 & Face Recognizer          & Computer vision-based               & Request-response & & LC, LTp, HCI \\ \cline{2-4} \cline{6-6} 
 & Database                  & General purpose        & Request-response &  & LT, LTp \\ \hline
\multirow{3}{*}{\shortstack{Industrial machinery\\ predictive maintenance \\application}} &
  Orchestrator &
  Time-series processing &
  Publish-subscribe & 
  \multirow{3}{*}{Aggregator} & LC, LTp
   \\ \cline{2-4} \cline{6-6} 
 & Emergency event detection  & Time-series processing & Request-response &  & LC, LTp, MCI \\ \cline{2-4} \cline{6-6} 
 & Missing data imputation & General purpose & Request-response &  & LC, LTp, MCI \\ \hline
\shortstack{Location retrieval \\application} &
  \shortstack{Location service with \\timed cache} &
  Simple non-resource-intensive &
  Request-response &
  Passthrough & LC, HTp
   \\ \hline
    \multicolumn{6}{|l|}{\begin{tabular}{l}LC: Latency Critical, LT: Latency Tolerant, HTp: High Throughput, MTp: Moderate Throughput, LTp: Low Throughput, HCI: High Compute Intensive, \\MCI: Moderate Compute Intensive, BI: Bandwidth Intensive\end{tabular}} \\ \hline
\end{tabular}%
}
\end{table*}

The generated dataset records data for three IoT applications typically deployed in edge environments:

\begin{figure}[th]
\centerline{\includegraphics[width=\columnwidth]{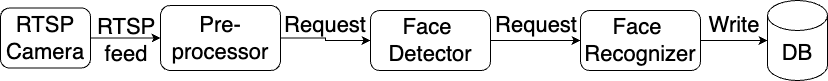}}
\caption{Face detection/recognition application}
\label{cv_app}
\end{figure}

\paragraph{Face detection/recognition} This application showcases the scenario of using computer vision for secure access control at a corporate office building. As illustrated in Figure \ref{cv_app}, it comprises four microservices: 1) preprocessor, 2) face detector, 3) face recognizer, and 4) database. Cameras at entry points (e.g., doors, gates) produce an RTSP stream. The preprocessor microservice reads from this video stream, performing resizing and grayscale conversion on the images and carrying out motion detection by thresholding the 
difference between consecutive frames. Upon detection of motion (i.e., when an employee approaches), the frame is sent to the face detector microservice. This microservice utilizes a Multi-Task Cascaded Convolutional Neural Network (MTCNN) to detect bounding boxes and landmark points of faces in the frame. Additionally, it conducts face alignment using affine transformation and supports multi-face alignment within a single frame. Aligned faces are then input to the face recognizer microservice. The face recognizer microservice employs a ResNet-50 model to extract features from the detected faces and calculates the cosine distance between these features and the features of the authorized personnel's faces to calculate the similarity between detected faces and faces of authorized personnel. Upon successful recognition, the timestamp is logged in the database - which is implemented as another microservice - marking the employee's entry or exit from the building.

\begin{figure}[th]
\centerline{\includegraphics[width=\columnwidth]{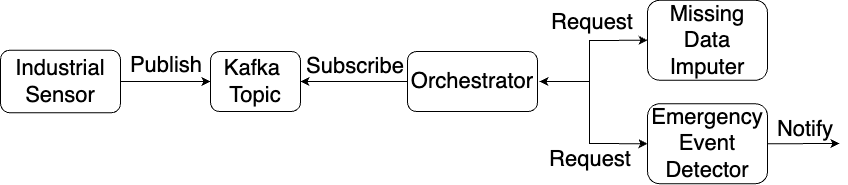}}
\caption{Industrial machinery predictive maintenance application}
\label{ts_app}
\end{figure}

\paragraph{Predictive maintenance for industrial machinery} In this application, IoT sensors that measure temperature, vibration, and pressure continuously generate multivariate time series data, which is then written to a Kafka topic. As shown in Figure \ref{ts_app}, an orchestrator microservice subscribes to this Kafka topic, reads the raw sensor data, and forms time-series windows, which are sent to the emergency event detector microservice. Before the windowed data is sent for emergency event detection, the orchestrator imputes any missing values by calling the missing data imputer microservice and also standardizes the data using standard score normalization. The emergency event detector uses an isolation forest model to detect whether a window of standardized time series data is anomalous by detecting patterns that deviate from the norm, such as overheating, excessive vibration, or mechanical stress. When an anomaly is detected, alerts are triggered to the maintenance teams. 

\begin{figure}[th]
\centerline{\includegraphics[width=0.75\columnwidth]{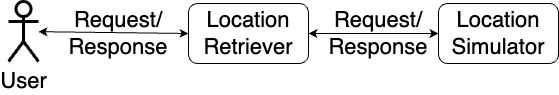}}
\caption{Location retrieval application}
\label{loc_app}
\end{figure}

\paragraph{Location retrieval} This application depicts the scenario of fleet management for logistics or delivery services. It facilitates tracking the real-time location of delivery vehicles in a fleet. As shown in Figure \ref{loc_app}, when a dispatcher or a customer requests the location of a specific vehicle, the request is directed to the location retriever microservice. This microservice checks its Least Recently Used (LRU) cache, and if the vehicle's location was recently queried, it provides the cached location for quick access. If the location is not in the cache, the microservice queries the vehicle's current location, updates the cache, and returns the most up-to-date information. In our implementation, the location simulator microservice is used to mock the location of the delivery vehicle by generating GPS locations in a trajectory.

As shown in Table \ref{tab:iot-app-properties}, the aforementioned applications span the properties of a wide range of IoT applications. For instance, the face detection/recognition application falls under Computer Vision (CV), the industrial machinery predictive maintenance application focuses on time-series processing, and the location retrieval application is a simple, non-resource-intensive application. Each application relies on different software architectures, including stream processing, request-response, and publish-subscribe, which are implemented using different communication protocols, including HTTP, Kafka, RTSP, and MySQL. The applications also employ different service composition patterns, such as chained, aggregator, and passthrough. Most importantly, the microservices in these applications have unique QoS and resource requirements.

The iAnomaly repository also contains the Python implementation for all three IoT applications. In addition, we have made the Docker images of the microservices accessible to the public\footnote{https://hub.docker.com/repository/docker/dtfernando/ianomaly}.

\subsection{Dataset Generation}
\label{subsec:implementation_details}

The applications presented in Section~\ref{subsec:iot_apps} were deployed in a Kubernetes cluster, where iAnomaly was responsible for the orchestration and automation of the data generation and collection process. Specifically, the physical environment consisted of ten VMs with heterogeneous computing and storage specifications created in the Melbourne Research Cloud\footnote{https://dashboard.cloud.unimelb.edu.au/} to emulate the worker nodes. In particular, two 2vCPU/8G VMs represented the IoT layer, four 2vCPU/8G VMs represented Fog level 1, three 4vCPU/16GB VMs represented Fog level 2, and one 8vCPU/32GB VM represented Fog level 3. The configuration of the emulated network was determined based on existing related research~\cite{pallewatta2019microservices}, with the following specifications: IoT layer $\rightarrow$ Fog level 1: 5ms/100Gbps, Fog level 1 $\rightarrow$ Fog level 2: 20ms/10Gbps, Fog level 2 $\rightarrow$ Fog level 3: 50ms/0.15Gbps, and 2ms bandwidth among nodes at the same level. The applications were assigned to the worker nodes by following the QoS-aware scheduling algorithm proposed by Pallewatta et al. in their research study \cite{pallewataa2022qos}. 

\begin{figure}[t]
\centerline{\includegraphics[width=0.8\columnwidth]{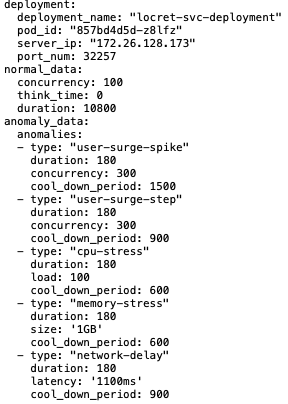}}
\caption{Dataset generation configuration YAML file of the location retrieval application}
\label{config_yaml}
\vspace{-15pt}
\end{figure}

\begin{figure}[t]
\centerline{\includegraphics[width=0.8\columnwidth]{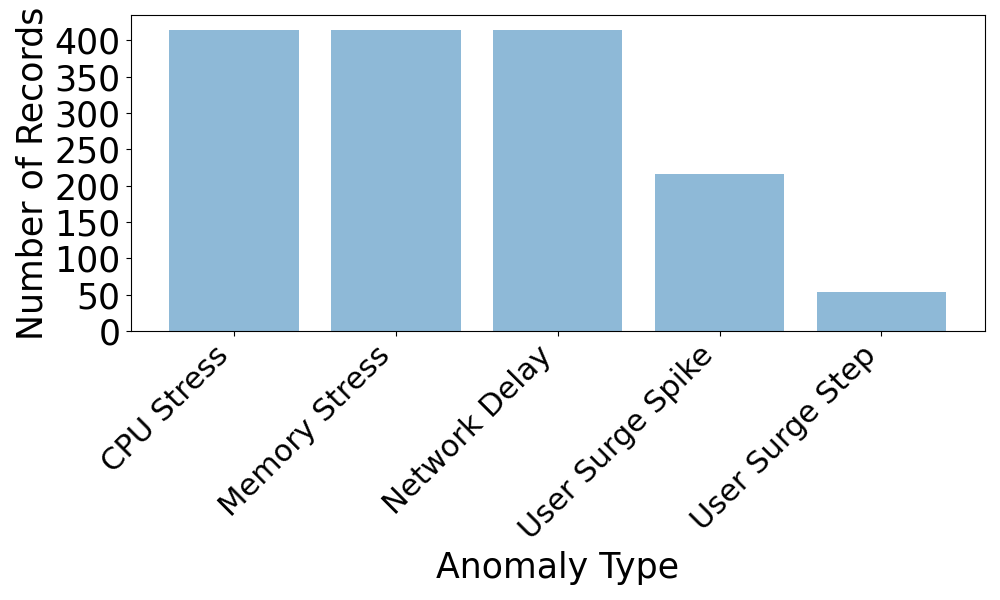}}
\caption{Distribution of records by anomaly type}
\label{anomaly_distribution}
\vspace{-15pt}
\end{figure}

Thereafter, normal and anomaly data were generated from each application by providing dataset generation configurations specified in the form of YAML files. For example, the configuration shown in Figure \ref{config_yaml} was used to generate data from the location retrieval application. This configuration instructs iAnomaly to generate and collect normal performance data from the location retriever microservice over a duration of three hours. Additionally, it specifies the injection of five types of anomalies—CPU hog, memory stress, user surge spike, user surge step, and network delay—over a total duration of two hours. Similar configurations were used to collect performance data from all applications. While five types of anomalies (two client-side and three server-side) were injected into the location retrieval application, only a subset of these anomalies was introduced into the other applications. This selection was based on the likelihood of each anomaly type occurring in real-world conditions for the respective applications. Figure \ref{anomaly_distribution} depicts the distribution of different types of anomalies across the dataset. It is also important to note that data for each application was collected independently to avoid anomalies caused by colocation.

Pixie's default granularity of 10 seconds was used when collecting data. For each application, data was collected across 12 metrics, covering key aspects of system and application performance: disk read and write throughput (total\_disk\_read\_throughput, total\_disk\_write\_throughput), memory usage (rss, vsize), CPU utilization (cpu\_usage), network activity (rx\_bytes\_per\_ns, tx\_bytes\_per\_ns), latency percentiles (latency\_p50, latency\_p90, latency\_p99), request throughput (request\_throughput), and error rate (errors\_per\_ns), collectively providing a comprehensive view of resource consumption, network efficiency, and service reliability. 

The final dataset comprises a total of 30240 records. Of these, 19260 records attribute to 54 hours of 
normal data, and 10980 records account for 31 hours of 
anomalous data. Within this dataset, there are 1512 records labelled as anomalous data points, resulting in an anomaly ratio of 5\%. This ratio is consistent with the anomaly ratio of other standard anomaly datasets, such as SMD (5.84\%) and ASD (4.61\%) \cite{li2021intermetric}, indicating that our dataset maintains a realistic anomaly density—an important characteristic of a high-quality anomaly dataset \cite{Renjie2022current}. The collected dataset is also made available at the iAnomaly repository. 

\subsection{Analysis of the Generated Dataset}
\label{subsec:dataset_analysis}

\begin{figure}[t]
\centerline{\includegraphics[width=\columnwidth]{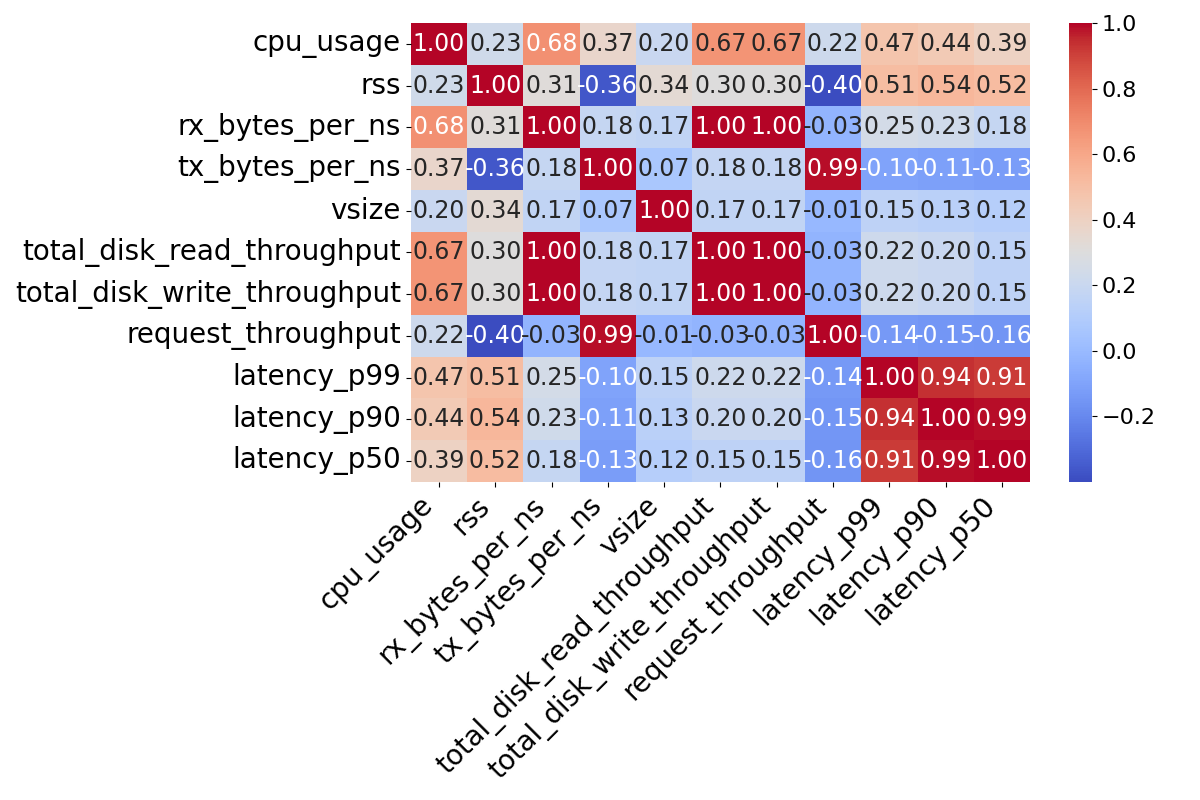}}
\caption{Collinearity among metrics in the generated dataset}
\label{colinearity_plot}
\vspace{-15pt}
\end{figure}

\begin{figure}[t]
\centerline{\includegraphics[width=\columnwidth]{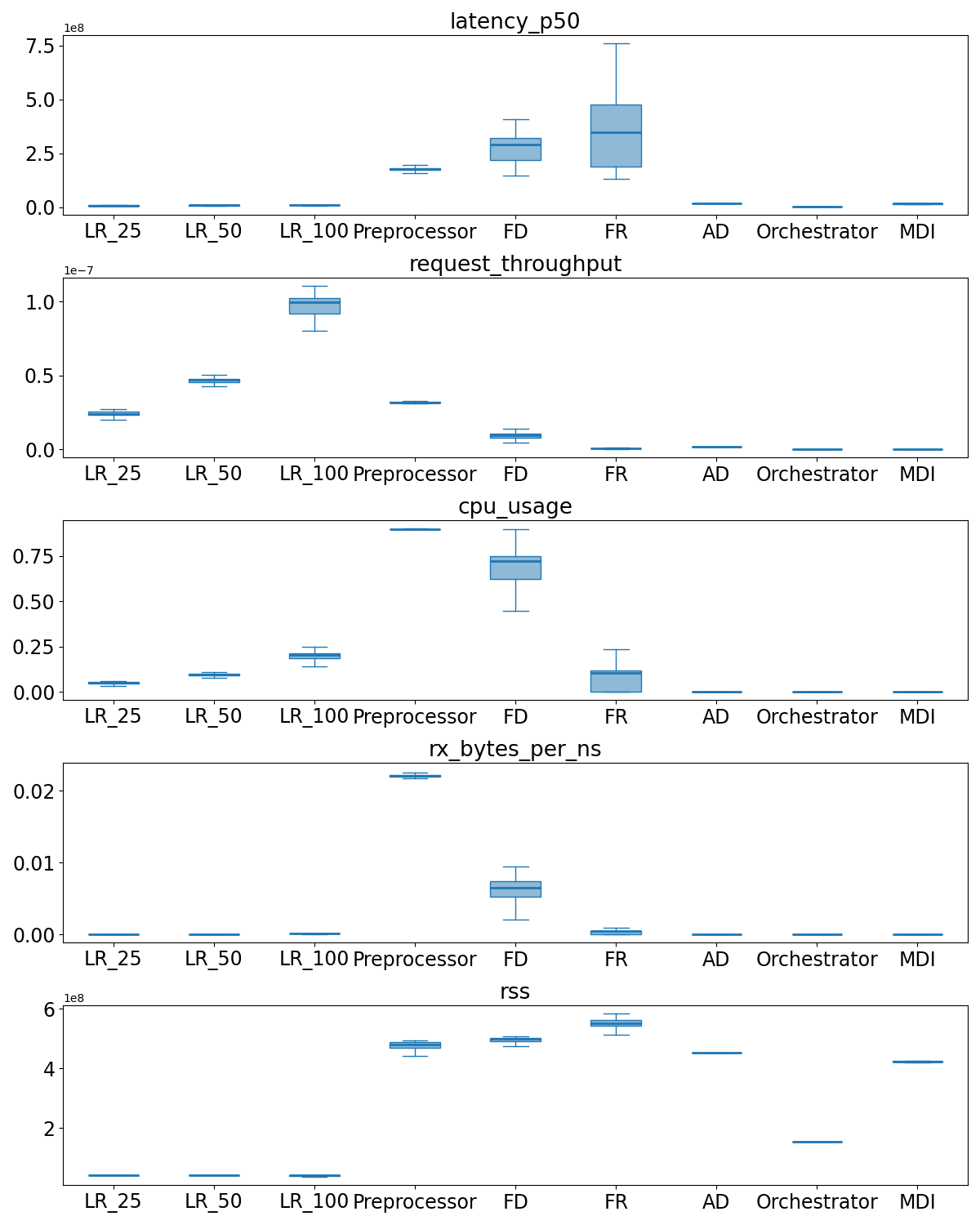}}
\caption{Distribution of normal data across shortlisted metrics}
\label{box_plots}
\vspace{-15pt}
\end{figure}

Figure \ref{colinearity_plot} illustrates the colinearity among the metrics in the generated dataset after excluding errors\_per\_ns metric, which contains all-zero values. The plot shows that inherently related groups of metrics, such as disk read/write throughputs, as well as latency percentiles, are highly correlated. In addition, metrics such as request\_throughput and tx\_bytes\_per\_ns, as well as disk read/write throughputs and rx\_bytes\_per\_ns, exhibit a strong positive correlation with each other. Outside of these groups, most other metrics do not show strong correlations with each other, indicating that each metric serves a unique purpose within the dataset. To simplify our analysis, we select a single metric from each identified group of correlated metrics to serve as a proxy for the others in the group. Based on this correlation analysis, we identify cpu\_usage, rss, rx\_bytes\_per\_ns, vsize, request\_throughput, and latency\_p50 as the subset of metrics with the lowest collinearity. Consequently, we will focus on these metrics for further analysis. 

Figure \ref{box_plots} illustrates the distribution of normal data for each shortlisted metric, focusing on three instances of the location retriever microservice and one instance of each of the other microservices. The three instances of the location retriever correspond to different deployments in regions with varying user populations. In the latency\_p50 subplot, it is evident that the preprocessor, face detector, and face recognizer microservices experience the highest latency. This increased latency is attributed to their highly compute-intensive (HCI) nature, which requires more time to process and respond to individual requests. In contrast, the other microservices demonstrate lower latency due to their relatively lower compute intensity. 

The second subplot represents the request\_throughput metric. Here, we can observe that the preprocessor and location retriever microservices exhibit high request throughput (HTp), while the face detector shows moderate throughput (MTp). The other microservices fall into the low throughput (LTp) category. These observations confirm the expected QoS properties of the microservices, as listed in Table \ref{tab:iot-app-properties}. The third subplot corresponds to the cpu\_usage metric. Despite being HTp, the location retriever microservices result in low CPU usage since they are not computationally intensive. The preprocessor microservice shows the highest CPU usage due to its HCI nature and high request throughput. The face detector microservice, while also HCI, has moderate throughput and, therefore, has the second-highest CPU usage. The rest of the microservices have a low CPU usage due to their LTp nature.

The rx\_bytes\_per\_ns subplot confirms the bandwidth-intensive (BI) nature of the preprocessor microservice. The final subplot, which corresponds to the rss metric, indicates that three computer vision microservices, together with the anomaly detector and the missing data imputer (which also utilize machine learning models for processing), have high rss values, demonstrating significant memory usage. By comparing these subplots, we can see that the diversity of the selected applications allows our collected dataset to effectively capture the variations in QoS and resource requirements of the microservices, as expected from an edge dataset. 

\begin{figure*}[t]
    \centering
    \subfigure[latency\_p50 of the preprocessor]{\includegraphics[width=.3\textwidth,
    height=3.25cm]{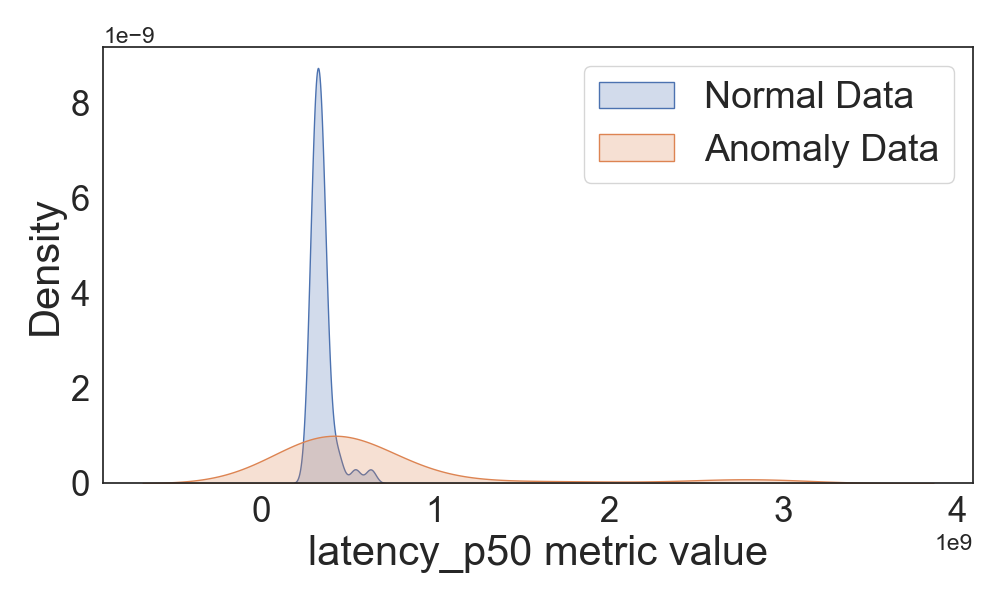}}
    \subfigure[cpu\_usage of the face detector]{\includegraphics[width=.3\textwidth, height=3.25cm]{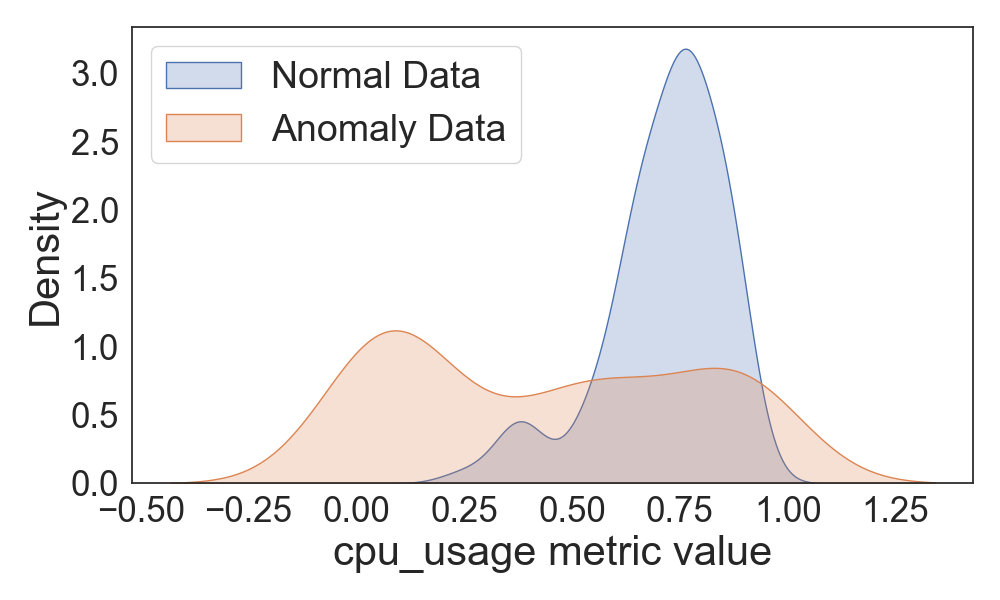}}
    \caption{PDFs for normal and anomalous data distributions of selected metrics}
    \label{pdfplots}
\end{figure*}

Figure \ref{pdfplots} contains the Probability Density Functions (PDFs) for the normal and anomalous data distributions of two randomly selected metrics from the datasets of the preprocessor and face detector microservices. Subfigure \ref{pdfplots}(a) corresponds to latency\_p50 of the preprocessor microservice while figure \ref{pdfplots}(b) corresponds to cpu\_usage of the face detector microservice. Both subplots illustrate that the anomalous data overlaps with the distribution of normal data. This overlap proves that the anomalies present in our dataset are non-trivial and not merely outliers. Renjie et al. \cite{Renjie2022current} have identified the presence of such non-trivial anomalies as a property of a good anomaly dataset.  

\begin{figure*}[t]
    \centering
    \subfigure[User surges using latency\_p50]{\includegraphics[width=.3\textwidth,
    height=3.25cm]{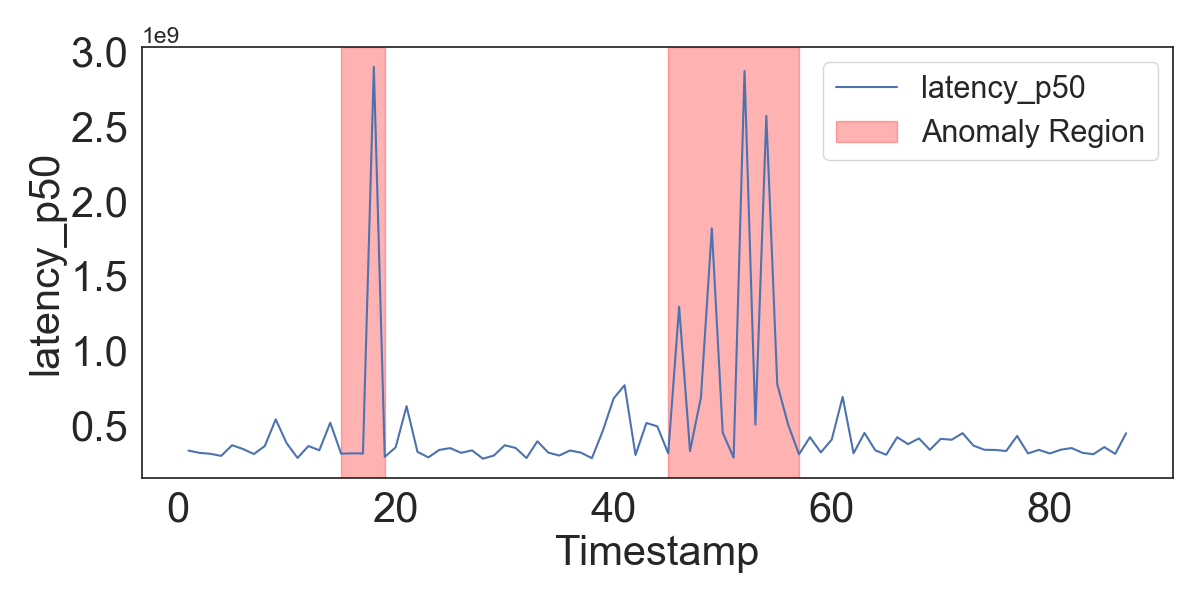}}
    \subfigure[Memory stress using rss]{\includegraphics[width=.3\textwidth, height=3.25cm]{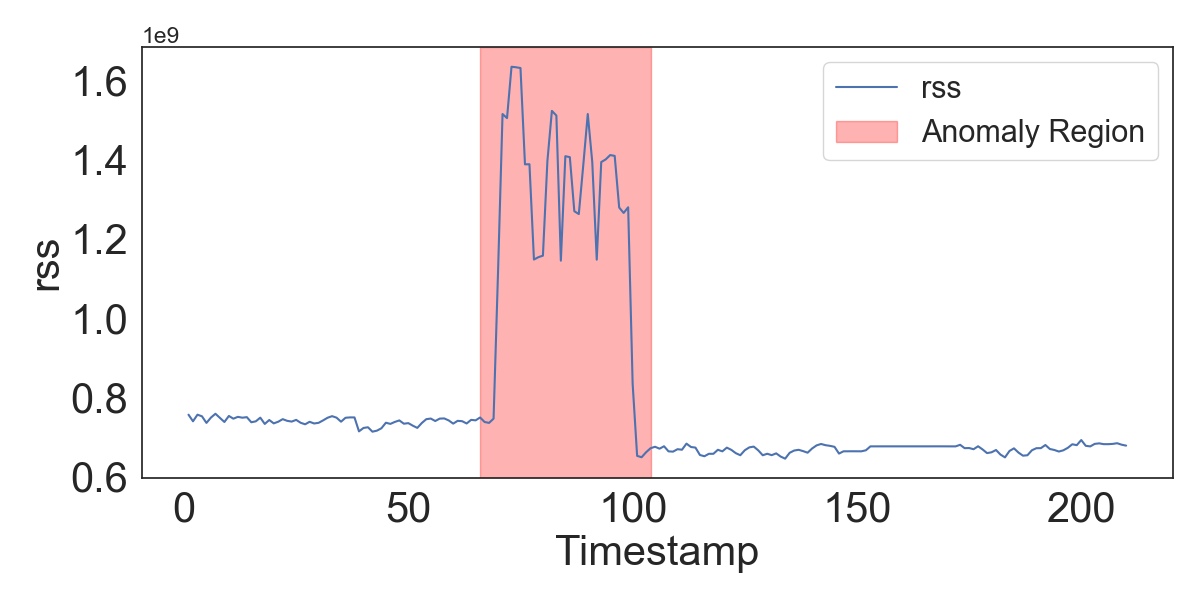}} 
    \subfigure[CPU stress using cpu\_usage]{\includegraphics[width=.3\textwidth, height=3.25cm]{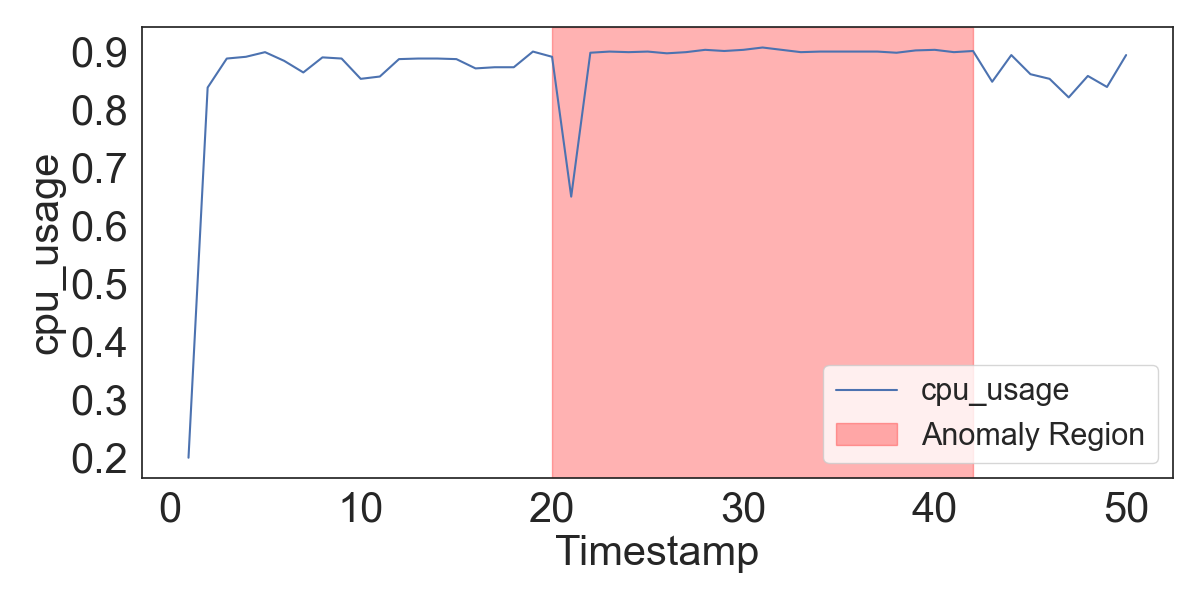}}
    \caption{Visualization of selected anomalies from the dataset}
    \label{anomaly_plots}
    \vspace{-15pt}
\end{figure*}

Furthermore, Figure \ref{anomaly_plots} visualizes a few selected anomalies from the dataset. While certain anomalies are easily noticeable using their respective metrics - for example, user surge anomalies are evident from the increase in the latency\_p50 metric (Figure \ref{anomaly_plots}(a)), and memory stress is apparent from the rss metric (Figure \ref{anomaly_plots}(b)) - some anomalies, such as CPU stress, cannot be detected simply by looking at the cpu\_usage metric (Figure \ref{anomaly_plots}(c)), especially when it occurs in compute-intensive microservices. During such scenarios, which are non-trivial to detect, algorithms that are capable of analyzing the higher-order relationships and behavior of several metrics are required to make an accurate detection.

Successful collection of the dataset was possible due to iAnomaly's use of an optimal set of open-source tools. In particular, leveraging Pixie as the monitoring tool allowed it to gather metric data from all three IoT applications, each using different communication protocols. In contrast, using a regular full-system emulator like iContinuum would only allow data collection from the location retrieval application, which uses HTTP for communication. Furthermore, iAnomaly's automated dataset generation capabilities led to an 87\% reduction in code lines compared to using a regular full-system emulator such as iContinuum during our dataset generation. Notably, iAnomaly completely eliminated the need for human intervention during the data collection process, requiring only 31 lines of configurations per microservice, while iContinuum needs 307 lines of code and significant human involvement to generate the same dataset. 

\section{Conclusions and Future Work}
\label{sec:conclusion}

Since existing research on performance anomaly detection in edge computing environments is limited due to the absence of publicly available edge performance anomaly datasets and due to the lack of accessibility of real edge setups to generate necessary data, we propose iAnomaly: a full-system emulator with performance anomaly dataset generation capabilities. Towards that, it is equipped with open-source tools such as Pixie for monitoring, Jmeter for normal workload generation, and client/sensor-side anomaly injection, as well as Chaos Mesh to introduce server-side anomalies. It also incorporates a dataset generation orchestrator to facilitate automatic data generation and collection based on user-defined configurations. 

As a case study, we generated a performance anomaly dataset using iAnomaly. It contains performance data for various microservice-based IoT applications with different QoS and resource requirements, injected with anomalies on both the client/sensor side and the server side. Analysis of this dataset showed that it represents the characteristics of real edge environments, and the anomalous data in the dataset meets the required standards for high-quality performance anomaly datasets. We have made this dataset available to the public. Additionally, we have released the iAnomaly toolkit for other researchers who may need to collect a more extensive dataset or conduct further anomaly detection research.

iAnomaly toolkit can easily be extended to collect trace data alongside metrics, which is particularly useful for research on root cause localization (RCL). Extending to such multi-source datasets is possible due to Pixie’s support. Researchers can further enhance this framework to conduct real-time experiments on anomaly-aware resource management. The toolkit and the released dataset are not limited to anomaly detection research. Normal data from the dataset, as well as normal data generated from the iAnomaly toolkit, can be used as foundational traces for experiments in other related research areas, such as resource scheduling and resource management. 

\bibliographystyle{IEEEtran}
	
\bibliography{reference}

\end{document}